*UQ@THU*

# Optimal monitoring location for risk tracking of geotechnical systems: theory and application to tunneling excavation risks

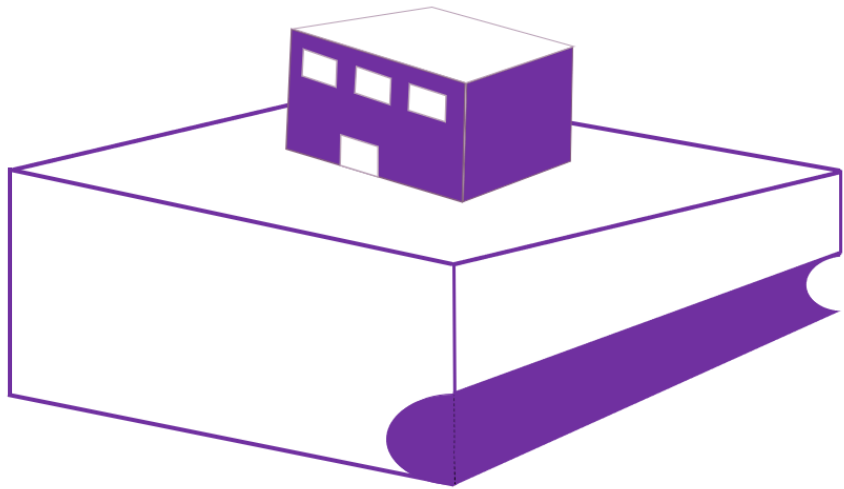

By: Dr. Zeyu Wang, Dr. Abdollah Shafieezadeh, Xiong Xiao,

Dr.Xiaowei Wang and Dr. Quanwang Li

Tsinghua university, School of civil engineering, Beijing, China

# Optimal monitoring location for risk tracking of geotechnical systems: theory and application to tunneling excavation risks

Zeyu Wang, Department of Civil Engineering, Tsinghua University, Beijing 100084, China
Abdollah Shafieezadeh[2], The Ohio State University, Columbus 43202, United States
Xiong Xiao, Department of Civil Engineering, Tsinghua University, Beijing 100084, China
Xiaowei Wang, Department of Bridge Engineering, Tongji University, Shanghai 200092, China
Quanwang Li, Department of Civil Engineering, Tsinghua University, Beijing 100084, China

**ABSTRACT**
The maturity of structural health monitoring technology brings ever-increasing opportunities for geotechnical structures and underground infrastructure systems to track the risk of structural failure, such as settlement-induced building damage, based on the monitored data. Reliability updating techniques can offer solutions to estimate the probability of failing to meet a prescribed objective using various types of information that are inclusive of equality and inequality. However, the update in reliability can be highly sensitive to monitoring location. Therefore, there may exist optimal locations in a system for monitoring that yield the maximum value for reliability updating. This paper proposes a computational framework for optimal monitoring location based on an innovative metric called sensitivity of information (SOI) that quantifies the relative change in unconditional and conditional reliability indexes. A state-of-the-practice case of risks posed by tunneling-induced settlement to buildings is explored in-depth to demonstrate and evaluate the computational efficiency of the proposed framework.

## 1. Introduction

Geotechnical structures and underground infrastructure systems are often subject to various forms of stressors such as subsidence that can threaten their functionality and safety. To capture those potentially unsafe conditions that may cause future catastrophic events, structural reliability analysis for geo-structures is indispensable. As sensor and, more broadly, monitoring technologies advance, enormously valuable information can be acquired without much effort. This brings new opportunities and challenges for risk analysis to integrate those data to reconsider the computational scheme of risk assessment. Grounded in the Bayesian updating theory, the emergence of reliability updating technique fills this gap by updating the probability of failure. In this context, let $F$ denote the failure event and $Z$ denote the observed information. Reliability updating aims to estimate the conditional probability of failure, $\Pr(F|Z)$, which can be formulated as (Straub, 2011),

$$\Pr(F|Z) = \frac{\Pr(F \cap Z)}{\Pr(Z)} \quad (1)$$

where $\Pr(F|Z)$ is the conditional probability of failure given information $Z$ (or the so-called posterior probability of failure) and $\Pr(F \cap Z)$ is the probability of the joint even $F \cap Z$. The information $Z$ can be generally classified into two groups that are inclusive of equality and inequality types. Computation of reliability updating with equality information is typically non-trivial through the conventional approaches such as surface integral (Straub, 2011) (Gollwitzer et al., 2006) and Bayesian networks (Straub, 2009; Straub Daniel and Der Kiureghian Armen, 2010a, 2010b). Fortunately, this computational challenge has been addressed by subtly introducing an auxiliary random variable to transform the equality information into an inequality one (Straub, 2011). However, the computation of $\Pr(F|Z)$ requires the probability of a joint event, $F \cap Z$, which is typically a very rare event. This probability can be estimated through subset simulation (*SS*) to improve the computational efficiency (Luque and Straub, 2016; Papaioannou and Straub, 2012). Moreover, by decomposing $\Pr(F \cap Z)$ into two more frequent probabilities $\Pr(Z)$ and $\Pr(F|Z)$ and training a surrogate model for the limit state function, metamodel-based approaches can facilitate fast estimation of $\Pr(F|Z)$ (Wang and Shafieezadeh, 2019a; Zhang et al., 2021).

Reliability updating has been recently applied in the field of geotechnical engineering for solving various types of practical problems. For example, field data and soil characteristics have been used to accurately estimate the reliability of a shallow foundation in a silty soil with spatially variable properties simulated via random fields (Papaioannou and Straub, 2017). Moreover, metamodeling techniques have been integrated to analyze the prior and posterior failure probabilities of a sheet pile wall in a dyke (van den Eijnden et al., 2021). This work demonstrated the computational capability of metamodel-based reliability updating in estimating $\Pr(F|Z)$. To ensure the safety of buildings in vicinity of a tunnel line, the settlement monitoring data at different locations were used to update the reliability of tunneling-induced settlement during excavation (Camós et al., 2016). This technique can better assist in risk management decisions if the ability of the planned tunneling line to satisfy the safety requirement can be checked in real-time through settlement monitoring. Analogous to this case, the deformation measurements of an excavation in sandy trench with a sheet pile retaining wall were also used to update the reliability of a construction site at its full excavation status (Papaioannou and Straub, 2012). This work can be viewed as a practical case for engineers in construction sites in avoiding catastrophic trench collapse. Additionally, to improve alarming system of a flood defense infrastructure, reliability updating together with head monitoring information were implemented in (Schweckendiek and Vrouwenvelder, 2013) to mitigate the risk of piping-induced levee failure in the presence of highly uncertain geohydrological properties. This work represents the potential capability of reliability updating in strengthening risk-informed warning systems against natural hazards. Moreover, reliability updating has been implemented in other applications including performance assessment of deteriorating reinforced concrete structures (Hackl and Kohler, 2016), structural inspection and repair of infrastructures (Yang and Frangopol, 2018), system identification (Lee and Song, 2016), life-cycle analysis (Straub Daniel and Der Kiureghian Armen, 2010b) and system reliability updating (Lee Young-Joo and Song Junho, 2014).

The reviewed literature showcases the high capability of reliability updating in successfully tracking the risk to geo-structures by incorporating the monitoring information within the existing computational scheme of reliability evaluation. However, very few of these studies paid attention to proper selection of monitoring location. In fact, the updated reliability can be very sensitive to the location where the monitoring information is obtained. Therefore, it can be inferred that there must exist optimal monitoring location, where the updated reliability can be utmost sensitive to the obtained information. To this end, this paper develops a method that efficiently determines the optimal monitoring location by introducing a novel metric called sensitivity of information (SOI) that measures the amplitude of the sensitivity at any location. Specifically, SOI is defined as the relative change in updated and prior reliability index, which facilitates the quantitative measurement of sensitivity of updated reliability index to the change of information at a specific location. To improve the computational efficiency of estimating $\Pr(F|Z)$, *SS* along with foregoing presented decomposition of $\Pr(F \cap Z)$ are integrated within the proposed computational framework. Moreover, the proposed SOI index subsequently parameterizes an objective function that is designed to find the optimal monitoring location by searching for its maxima based on a surrogate-assisted optimization. To examine the efficacy and computational efficiency, a state-of-the-practical case of tunneling-induced settlement to building damage is investigated.

The rest of this article is mainly organized in 5 sections. Section 2 briefly introduces the concept of reliability updating. Section 3 presents the proposed SOI index together with the framework for determining the optimal monitoring location. Subsequently, section 4 presents the procedures of analyzing SOI and exploring the optimal settlement monitoring location for a practical case that investigates the risk posed by tunneling-induced settlements. Conclusion remarks are drawn in section 5.

## 2 Reliability updating with equality information

Generally, the main difference between reliability analysis and updating lies in whether the observational information is available or not. Reliability analysis focuses on the computation of unconditional probability of failure $\Pr(F)$ while reliability updating estimates the conditional probability of failure $\Pr(F|Z)$. Let $g(X)$ denote the performance function, the response of which determines the condition of the system: $g(X) \leq 0$ indicates failure and $g(X) > 0$ means safe state; the boundary region where $g(X) = 0$ is called the limit state. Thus, the unconditional probability of failure can be defined as:

$$\Pr(F) = \Pr(g(X) \leq 0) \tag{2}$$

Methods for computing $\Pr(F)$ include but are not limited to: crude Monte-Carlo simulation (*MCS*) (Fishman, 2013; Rubinstein and Kroese, 2016), first- or second-order reliability analysis method (Kiureghian Armen Der and Stefano Mario De, 1991; Rackwitz and Flessler, 1978), importance sampling (IS) (Au and Beck, 1999; Hohenbichler and Rackwitz, 1988), *SS* (Au and Beck, 2003, 1999; Papaioannou et al., 2015) and surrogate-based methods (Echard et al., 2011; Wang and Shafieezadeh, 2020, 2019b, 2019c). As Eq. (1) shows, the estimate of $\Pr(F|Z)$ needs to compute $\Pr(Z)$ and $\Pr(F \cap Z)$. According to (Straub, 2011), the probability of information $\Pr(Z)$ can be computed as follows,

$$\Pr(Z) = \int_{\theta \in \Omega_\theta} \Pr(Z|\Theta(X) = \theta) f(\theta) d\theta \tag{3}$$

where $X$ denotes the vector of random variables, $\Theta(X)$ denotes a function parameterized by $X$ with the realization notation $\theta$, that can represent the uncertainty of the system characteristic, $\Theta_s(X)$, or the external loadings, $\Theta_e(X)$. Moreover, $f(\cdot)$ represents the probability density function (PDF) and $\Omega_\theta$ is the probabilistic space of $\Theta(X)$. In this context, the probability of the joint event $\Pr(F \cap Z)$ can be derived as,

$$\Pr(F \cap Z) = \int_{\theta \in \Omega_\theta} \Pr(F|\Theta(X) = \theta) \Pr(Z|\Theta(X) = \theta) f(\theta) d\theta \tag{4}$$

The likelihood of $\theta$ given the information Z then can be formulated as,

$$L(\theta|Z) = \Pr(Z|\Theta(X) = \theta) \tag{5}$$

where $L(\cdot)$ denotes the likelihood function. To further elaborate the above equation, let's consider a case where Z represents the measurements $m_s$ of a property of the system (i.e., $\Theta_s(X)$) with measurement error $\varepsilon_s$ and $m_e$ of the external loadings (i.e., $\Theta_e$) with measurement error $\varepsilon_e$. Note that $m_s$ is different from $\theta_s$. $m_s$ represents the information while $\theta_s$ is a stochastic realization of $\Theta_s$. Thus, if the two information pieces are mutually independent, the likelihood function can be represented as,

$$\begin{aligned} L(\theta|Z) &= L(\Theta_s(X) = \theta_s|Z = m_s) \cdot L(\Theta_e(X) = \theta_e|Z = m_e) \\ &= f_{\varepsilon_s}[m_s - \theta_s] \cdot f_{\varepsilon_e}[m_e - \theta_e] \end{aligned} \tag{6}$$

with $f_{\varepsilon_s}$ and $f_{\varepsilon_e}$ being the probability density functions (PDFs) of $\varepsilon_s$ and $\varepsilon_e$, respectively. The difference between $L(\Theta_s(X) = \theta_s|Z = m_s)$ and $L(\Theta_e(X) = \theta_e|Z = m_e)$ lies in whether the investigation of the performance function is needed or not. Moreover, $L(X|Z)$ and $L(X = x|Z)$ are two special cases of $L(\theta|Z)$ by denoting that $\Theta(X) = X$ and $\Theta(X) = x$. $L(X = x|Z)$ and $L(X|Z)$ are represented in the form of $L(x)$ and $L(X)$ from this point forward. Note that the following identity holds true for any likelihood functions:

$$L(x) = \frac{1}{c} \Pr\{U - \Phi^{-1}[cL(x)] \leq 0\} \tag{7}$$

where $c$ is a constant satisfying $0 \leq cL(x) \leq 1$, $\Phi^{-1}$ denotes the inverse standard normal cumulative distribution function, and $U$ represents a standard normal variable.

By reformulation the equality information into inequality one, $\Pr(Z)$ can be estimated by introducing the auxiliary random variable, $U$, and define an augmented Limit State Function (LSF),

$$\Pr(Z) = \alpha P(h(U, \mathbf{X}) \leq 0) \tag{8}$$

where $\alpha$ is a trivially computational constant, $h(U, \mathbf{X})$ is the augmented limit state function with an auxiliary standard uniform random variable, $P$ (Straub, 2011),

$$h(U, \mathbf{X}) = U - \Phi^{-1}[cL(\mathbf{X})] \tag{9}$$

Similarly, $\Pr(F \cap Z)$ can be computed by defining a limit state function that takes the maximum value of $g(\mathbf{X})$ and $h(U, \mathbf{X})$,

$$\Pr(F \cap Z) = \alpha P(\max[g(\mathbf{X}), h(P, \mathbf{X})] \leq 0) \tag{10}$$

Derivation of Eq. (10) is not elaborated in this paper for the sake of brevity. Detailed derivation can be found in (Straub, 2011). Note that $U$ is not necessarily a standard normal random variable, it can be as simple as a standard uniform distributed random variable. Therefore, one can rewrite Eq. (9) as,

$$h(U, \mathbf{X}) = P - cL(\mathbf{X}) \tag{11}$$

However, the adoption of standard normal random variable can improve the smoothness of the responses of the function. To increase the readability, the computational scheme based on Eq. (9) is used throughout the paper. Combining Eq. (8) and (10), the conditional probability of failure can be obtained by canceling out the constant $\alpha$,

$$\Pr(F|Z) = \frac{P(J(U, \mathbf{X}) \leq 0)}{P(h(U, \mathbf{X}) \leq 0)} \tag{12}$$

where $J(U, \mathbf{X}) = \max[g(\mathbf{X}), h(U, \mathbf{X})]$. Eq. (12) enables fast reliability updating by solving two structural reliability problems. Typically, the numerator in Eq. (12) is very rare, which requires powerful structural reliability methods such as subset simulation that has the capability of estimating the probability of very rare events (Luque and Straub, 2016; Papaioannou and Straub, 2012). In the following context, an efficient and robust approach for the estimation of $\Pr(F|Z)$ is presented, which facilities the localization of optimal monitoring location.

### 3 Optimal monitoring location analysis with SOI
Data measured at different locations of structures and infrastructure systems may have distinct impacts on the updated reliability. To precisely quantify this difference, a concept of sensitivity of information for the updated reliability is proposed in this paper. Moreover, the proposed concept can be further leveraged to identify the optimal monitoring location that makes the most significant contribution to the change of updated reliability. In this section, the concepts of sensitivity of information (SOI) along with the computational details are elaborated. By maximizing the objective function involving SOI, the optimal monitoring location can be derived with the goal of increasing the efficacy of risk tracing for structures and infrastructure systems.

#### 3.1 Sensitivity of information analysis for reliability updating

In practical engineering, acquiring information is typically costly; therefore, engineers should prudently select a worthwhile location for structural monitoring and diagnosis. However, information collected in some locations has very neglectable impact on the change of updated reliability. On the other hand, the updated reliability is very sensitive to the information stemming from very valuable locations. Therefore, the level of sensitivity of updated reliability to the change of information should be mathematically quantified. In this paper, the foregoing concept is denoted as sensitivity of information. Let $\Pr(F|Z = z, L = l)$ represent the conditional probability of failure given the specific equality information $z$ and the monitoring location $l$, which can be calculated based on Eq. (12). The relative difference of the conditional reliability index, $\beta_{post}$, compared to the unconditional reliability index, $\beta_{prior}$, can be calculated as,

$$r_{up}(Z = z, L = l) = \left|\frac{\beta_{post}}{\beta_{prior}} - 1\right|$$
$$= \left|\frac{-\Phi^{-1}[\Pr(F|Z = z, L = l)]}{-\Phi^{-1}[\Pr(F)]} - 1\right| = \left|\frac{\Phi^{-1}[\Pr(F|Z = z, L = l)]}{\Phi^{-1}[\Pr(F)]} - 1\right| \quad (13)$$

where $r_{up}$ denotes the relative change in reliability. However, the information $z$ is unknown before it is measured at the location $l$. In fact, $z$ can be any number from $-\infty$ to $+\infty$ without any prior knowledge. However, some ranges can be unrealistic. Therefore, it is assumed that $z$ is uniformly distributed over the interval $[Z_{lob}, Z_{upb}]$, where $Z_{lob}$ and $Z_{upb}$ represent the lower and upper bounds of possible information which can be determined by engineering judgement. Therefore, the expected value of $r_{up}(Z, L = l)$ can be adopted to reflect the magnitude of $r_{up}(Z, L = l)$. In this paper, the sensitivity of information at location $l$ is computed as,

$$\text{SOI}(L = l) = \int_{-\infty}^{+\infty} r_{up}(Z = z, L = l) f_u(z) dz \approx \frac{1}{Z_{upb} - Z_{lob}} \int_{Z_{lob}}^{Z_{upb}} r_{up}(Z = z, L = l) dz \quad (14)$$

It can be inferred from Eq. (14) that SOI varies with location $L$. Moreover, a monitoring location with a large SOI tends to have significant impact on the change of reliability index while monitoring location with a small SOI indicates that the monitoring action is not valuable. The computation of Eq. (14) involves an operation of integral, which requires numerical discretization. Hence, the computational complexity depends on the scheme of such numerical discretization. Assume that the integral space is discretized into $n_{dis}$ pieces. Subsequently, Eq. (14) can be calculated as,

$$\text{SOI}(L = l) \approx \frac{1}{Z_{upb} - Z_{lob}} \sum_{i=1}^{n_{dis}} r_{up}(Z = z_i, L = l) \Delta_z \quad (15)$$

where $z_i = (2i - 1)\Delta_z$ is the point centered at the integral pieces and $\Delta_z = (Z_{upb} - Z_{lob})/n_{dis}$. Eq. (14) needs to investigate the estimate of reliability updating $n_{dis}$ times, which is computationally very intensive and not practical. Concerning this issue, the computation of $r_{up}(Z = z, L = l)$ in Eq. (13) needs to be optimized.

### 3.2 Computational details of SOI
The computation of $r_{up}(Z = z, L = l)$ needs to investigate the estimates of $\Pr(F)$ and $\Pr(F|Z)$ for $n_d$ times. These probabilities can be possibly rare for some cases. To enhance the computational efficiency and robustness of the estimates of $\Pr(F)$ and $\Pr(F|Z)$, *SS* along with a strategy of decomposing $\Pr(F \cap Z)$ into $\Pr(Z|F) \cdot \Pr(F)$ is utilized in this paper. Therefore, the following equation is represented to estimate $\Pr(F|Z)$,

$$\Pr(F|Z) = \frac{\Pr(Z|F) \cdot \Pr(F)}{\Pr(Z)} \tag{16}$$

Eq. (16) optimizes the computation of Eq. (12) by decomposing $\Pr(F \cap Z)$ into $\Pr(Z|F)$ and $\Pr(F)$ via Bayes' theorem. This strategy completely avoids the computation of the probability of the rare event of $\Pr(F \cap Z)$. Integrating with *SS*, Eq. (16) can be rewritten as,

$$\Pr(F|Z) = \frac{\Pr(Z|F)}{\Pr(Z)} P\left(\bigcap_{i=1}^{m} F_i\right) = \frac{\Pr(Z|F)}{\Pr(Z)} P(F_1) \prod_{i=1}^{m-1} P(F_{i+1}|F_i) \tag{17}$$

where $F_i$ denotes the intermediate failure event of $g(X)$, $m$ denotes the number of subsets and $F_m$ is the target failure event. Given that $F = F_m$, Eq. (17) can be further simplified as,

$$\Pr(F|Z) = \frac{\Pr(Z|F_m)}{\Pr(Z)} P(F_1) \prod_{i=1}^{m-1} P(F_{i+1}|F_i) \tag{18}$$

This indicates that the computation of $\Pr(F|Z)$ only relies on the estimates of $\Pr(Z)$ and $\Pr(Z|F_m)$ once the estimate of $\Pr(Z)$ is completed through *SS*. For different information, the estimate of $\Pr(F|Z)$ can be as simple as reevaluating $\Pr(Z)$ and $\Pr(Z|F_m)$ based on the samples remained in the last target subset. However, we often encounter the situation where $\Pr(F|Z)$ cannot be estimated with sufficient accuracy due to the insufficient samples. This inaccuracy can lead to the associated inaccurate computation of $SOI(L = l)$. To overcome this limitation, samples generated through Markov Chain Monte Carlo simulation (*MCMC*) in each subset $S^i, i = 1,2,...,m$ should be sufficient so that $COV_{P_{F|Z}}$ is smaller than $COV_{thr}$, where $COV_{P_{F|Z}}$ and $COV_{thr}$ denote the coefficient of variation (COV) of $\Pr(F|Z)$ and the prescribed threshold, respectively. Toward this goal, the number of intermediate failure samples for *SS* is redefined in an adaptive way to facilitate the robust estimation of $\Pr(F|Z)$. Therefore, procedures for estimating $\Pr(F|Z)$ through the adaptive adjustment of $N_{ss}$ is summarized in the following procedures:

- **Step 1**: Define initial parameters $COV_{thr}$, $N_{ss}$ and $p_0$ for *SS*. In this paper, the UQLab toolbox with *Reliability* module in MATLAB® software is used. Other sets for performing *SS* follows the default settings in UQLab (Marelli et al., n.d.; "UQLab input manual," 2017).

- **Step 2**: Perform *SS* and record the computational results such as $P_F$, $t_i$s, $COV_{P_F}$ and $S^i$ etc. In this step, the proposal distribution for *MCMC* is selected to be uniform. Moreover, a conceptual illustration of this computation for a 2D problem is presented in Fig 1. For this step, all the failure samples are kept for the sake of computing $\Pr(Z|F)$ in step 4.

- **Step 3:** Estimate $\Pr(Z)$ with the following limit state function,

$$h_1(p, x) = P - c_1 L(X) \tag{19}$$

In most cases, $\Pr(Z)$ can be estimated through MCS if $L(x)$ does not need to investigate sophisticated model $g(x)$. Otherwise, $\Pr(Z)$ can be estimated through *SS*.

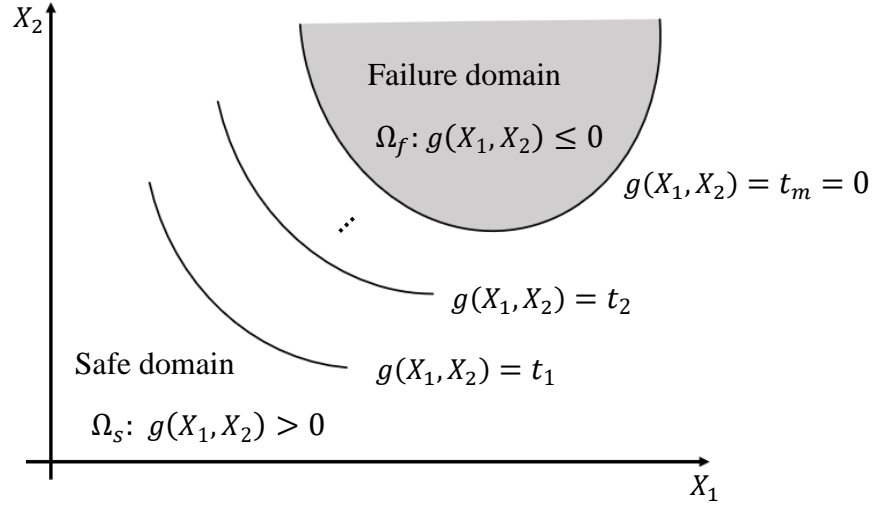

**Fig 1.** Illustration of SS for a 2D example with safe and failure domains and the limit state $g(X_1, X_2) = 0$

- **Step 4:** Estimate $\Pr(Z|F)$ based on the kept failure samples in step 2 with the following limit state function,

$$h_2(P, \mathbf{X}) = P - c_2 L(\mathbf{X}) \tag{20}$$

- **Step 5:** Check if $\text{COV}_{P_{F|Z}} \leq \text{COV}_{thr}$. Go to Step 6 if satisfied; otherwise, reset $N_{ss} = N_{ss}^{last} + \Delta N_{ss}$ and go back to Step 2, where $N_{ss}^{last}$ denotes the number of intermediate failure samples in each subset in the last iteration.

- **Step 6:** Output $\Pr(F)$ and $\Pr(F|Z)$.

Essentially, step 5 investigates the computation of $\text{COV}_{P_{F|Z}}$ which impacts the computational robustness of the updated reliability. Let $P_F$, $P_Z$, $P_{Z|F}$ and $P_{F|Z}$ denote $\Pr(F)$, $\Pr(Z)$, $\Pr(Z|F)$ and $\Pr(F|Z)$ for the sake of readability of this manuscript. To this end, $\text{COV}_{P_{F|Z}}$ is computed in the following context. In virtue of the equality $\text{Var}(AB) = [E(A)]^2 \text{Var}(B) + [E(B)]^2 \text{Var}(A) + \text{Var}(A)\text{Var}(B)$, where $A$ and $B$ are two mutually independent random variables, the following equation holds true,

$$\text{COV}_{P_{F|Z}} = \frac{\text{Var}\left(P_{F \cap Z} \frac{1}{P_Z}\right)^{\frac{1}{2}}}{E\left(P_{F \cap Z} \frac{1}{P_Z}\right)}$$

$$= \frac{\left[[E(P_{F \cap Z})]^2 \text{Var}\left(\frac{1}{P_Z}\right) + \left[E\left(\frac{1}{P_Z}\right)\right]^2 \text{Var}(P_{F \cap Z}) + \text{Var}(P_{F \cap Z})\text{Var}\left(\frac{1}{P_Z}\right)\right]^{\frac{1}{2}}}{E(P_{Z|F}) E(P_F) E\left(\frac{1}{P_Z}\right)} \tag{21}$$

where $P_F$, $P_{F \cap Z}$ and $P_{Z|F}$ denote $\Pr(F)$, $\Pr(F \cap Z)$ and $\Pr(Z|F)$, $E(\cdot)$ and $\text{Var}(\cdot)$ represent the operations of mean and variance. Moreover, $E(P_{F \cap Z}) = E(P_{Z|F}) E(P_F)$. The computation of $\text{Var}(P_{F \cap Z})$, $E\left(\frac{1}{P_Z}\right)$ and $\text{Var}\left(\frac{1}{P_Z}\right)$ is elaborated next. First, $\text{Var}(P_{F \cap Z})$ can be estimated according to the following equation,

$$\mathrm{Var}(P_{F\cap Z}) = \mathrm{Var}(P_F P_{Z|F})$$
$$= [\mathrm{E}(P_F)]^2 \mathrm{Var}(P_{Z|F}) + [\mathrm{E}(P_{Z|F})]^2 \mathrm{Var}(P_F) + \mathrm{Var}(P_F)\mathrm{Var}(P_{Z|F}) \qquad (22)$$

If $N_{ss}$ is sufficiently large, the following equation holds true,

$$\lim_{N_{ss} \to 0} \mathrm{E}(P_F) \cong \tilde{P}_F \qquad (23)$$

where $\tilde{P}_F$ denotes the ground truth of the unconditional probability of failure. The variance of $P_F$ can be correspondingly calculated as,

$$\mathrm{Var}(P_F) = \mathrm{COV}_{P_F}^2 [\mathrm{E}(P_F)]^2 \qquad (24)$$

and the COV of $P_f$, $\mathrm{COV}_{P_F}$, is calculated as,

$$\mathrm{COV}_{P_F} = \sqrt{\sum_{i=1}^{m} \mathrm{COV}_{\hat{p}_i}^2}, \text{ for } i = 2,3,\ldots,m \qquad (25)$$

Generally, the COV of each $\hat{p}_i$ can be estimated as,

$$\mathrm{COV}_{\hat{p}_1} = \sqrt{\frac{1-P_i}{P_i N}}, \qquad \text{for } i = 1 \qquad (26)$$

and

$$\mathrm{COV}_{\hat{p}_i} = \sqrt{\frac{1-P_i}{P_i N}(1+\gamma_i)}, \qquad \text{for } i = 2,3,\ldots,m \qquad (27)$$

where $\gamma_i$ is a computational index that can be determined as,

$$\gamma_i = 2 \sum_{k=1}^{N/N_c - 1} \left(1 - \frac{kN_c}{N}\right) \rho_i(k) \qquad (28)$$

where $\rho_i(k)$ denotes the correlation coefficient at lag $k$ of the stationary sequence $\{I_{j,k}^{\{i\}}: k = 1, \ldots, N/N_c\}$, which can be calculated as,

$$\rho_i(k) = R_i(k)/R_i(0) \qquad (27)$$

The covariance sequence $\{R_i(k): i = 0, \ldots, N/N_c - 1\}$ can be estimated based on Markov chain samples as follows,

$$R_i(k) \cong \left(\frac{1}{N_{ss} - kN_c} \sum_{j=1}^{N_c} \sum_{i=1}^{N_{ss}/N_c - k} I_{j,l}^{\{i\}} I_{j,l+k}^{\{i\}}\right) - p_i^2 \qquad (28)$$

where $N_c$ denotes the number of Markov chains and $I_{j,l}^{\{i\}}$ denotes the failure indicator for the $k$th sample in the $j$th Markov chain simulation level $i$. Moreover, the mean and the variance of $P_{Z|F}$ are estimated as,

$$\mathrm{E}(P_{Z|F}) = P_{Z|F} \tag{29}$$

and

$$\mathrm{Var}(P_{Z|F}) = NP_{Z|F}(1 - P_{Z|F}) \tag{30}$$

Therefore, $\mathrm{Var}(P_{F \cap Z})$ can be obtained by combining Eq. (20) ~ (30). Moreover, $\mathrm{E}\left(\frac{1}{P_Z}\right)$ and $\mathrm{Var}\left(\frac{1}{P_Z}\right)$ can be estimated through numerical simulation by taking the reciprocal of a normal random variable (The Central Limit Theorem) after its COV and mean are acquired. Importantly, the computation of COV of $P_Z$ depends on the type of reliability method (i.e., MCS or *SS*). One should note that the proposed approach overperforms those approaches that rely on the limit state function of joint event (i.e., $J(U, X)$ defined in Eq. (12)) (Luque and Straub, 2016; Papaioannou and Straub, 2012). Different from the computational scheme that needs to estimate $\Pr(F \cap Z)$, $\Pr(Z|F)$ only focuses on the failure domain of the performance function, which completely avoids unnecessary computational efforts of estimating $\Pr(Z|F)$.

### 3.3 Analysis of optimal monitoring location

The metric SOI can be leveraged to derive the optimal monitoring location that has the most significant impact on the change of updated reliability index. Generally, the optimal monitoring location can be identified according to the following equation,

$$l^* = \arg\max_{l \in \Gamma} \mathrm{SOI}(L = l) \tag{35}$$

where $l^*$ denotes the vector of optimal monitoring location ($l^*$ is not bold as $l^*$ if it denotes one location), $\Gamma$ represents the domain of the global feasible monitoring locations and $\mathrm{SOI}(L = l)$ represents the sensitivity of information at location $l$. However, the optimization problem represented in Eq. (35) can be computationally prohibitive due to the complex topology of monitoring location with large dimension or discretization. To further interpret this point, let $N_T$ denote the total number of discretized mapping points, it can be calculated as,

$$N_T = \prod_{i=1}^{N_{dim}} N_i^d \tag{36}$$

where $N_{dim}$ is the number of the dimension of $\Gamma$ and $N_i^d$ represents the number of discretized points in the $i^{th}$ dimension. For example, $N_T$ can be as large as $10^6$ for topology with three dimensions if it is discretized into 100 pieces in each dimension. The optimization defined in Eq. (35) becomes computationally intractable if SOI of all these discretized points are calculated. To efficiently solve the optimization problem in Eq. (35), a surrogate model-based optimization solution is adopted to find $l^*$. In this paper, the Kriging surrogate model with noisy responses is adopted to tackle the inconsistent estimate of SOI presented in section 3.2. Based on Kriging surrogate model with noisy response, SOI for each discretized sample $l$ can be represented as:

$$\widehat{\mathrm{SOI}}(l) = F(l, \beta) + \Psi(l) + \epsilon_k = f^T(l)\beta + \Psi(l) + \epsilon_k, \tag{37}$$

where $\widehat{\mathrm{SOI}}(l)$ denotes the value of SOI at $L = l$ estimated through the Kriging surrogate model, $\Psi(l)$ denotes the Gaussian process, $\epsilon_k$ is the additive noise of response which follows a zero-mean Gaussian distribution with covariance matrix $\Sigma_n$, and $F(l, \beta)$ is the so-called regression basis denoting the Kriging trend, which can be a constant, a polynomial term or any mathematical form. Moreover, $f(l)$ is the vector of Kriging basis and $\beta$ is the vector of regression coefficients. Specifically, $f^T(l)\beta$ often takes the form of ordinary ($\beta_0$), linear ($\beta_0+\sum_{n=1}^{N_{dim}}[\beta]_n[x]_n$) or quadratic ($\beta_0+\sum_{n=1}^{N_{dim}}[\beta]_n[x]_n+\sum_{n=1}^{N_{dim}}\sum_{k=n}^{N_{dim}}[\beta]_{nk}[x]_n[x]_k$), where $[\beta]_n$ and $[x]_n$ denote the n$^{th}$ component of $\boldsymbol{\beta}$ and $\boldsymbol{l}$. Moreover, $\Psi(\boldsymbol{l})$ has a zero mean and a covariance matrix between two points, $\boldsymbol{l}_i$ and $\boldsymbol{l}_j$:

$$\text{COV}\left(\Psi(\boldsymbol{l}_i), \Psi(\boldsymbol{l}_j)\right) = \sigma^2 R(\boldsymbol{l}_i, \boldsymbol{l}_j; \boldsymbol{\theta}), \tag{38}$$

where $\sigma^2$ is the process variance or the generalized mean square error from the regression part and $R(\boldsymbol{l}_i, \boldsymbol{l}_j; \boldsymbol{\theta})$ is the correlation function or the kernel function representing the correlation function of the process with hyper-parameter $\boldsymbol{\theta}$. Multiple types of correlation functions are available for Kriging models including linear, exponential, Gaussian, Matérn models, among others ("UQLab Kriging (Gaussian process modelling) manual," n.d.). In this paper, the Gaussian kernel function is implemented:

$$R(\boldsymbol{l}_i, \boldsymbol{l}_j; \boldsymbol{\theta}) = \prod_{n=1}^{N_{dim}} \exp\left(-[\theta]_n \left([\boldsymbol{l}_i]_n - [\boldsymbol{l}_j]_n\right)^2\right), \tag{39}$$

where $[\boldsymbol{l}_i]_n$ is the n$^{th}$ component of the realization $\boldsymbol{l}_i$, $\boldsymbol{\theta}$ denotes the hyper-parameter that can be estimated via maximum likelihood estimation (MLE) or cross-validation ("UQLab Kriging (Gaussian process modelling) manual," n.d.). It is shown that the Kriging prediction is very sensitive to the value of $\boldsymbol{\theta}$ (Kaymaz, 2005; Wang et al., 2017; Wen et al., 2016). In this article, the optimal hyper-parameter $\boldsymbol{\theta}^*$ is searched through MLE:

$$\boldsymbol{\theta}^* = \underset{\boldsymbol{\theta} \in \mathbb{R}}{\text{argmin}} \frac{1}{2}\left[\log\left(\det\left(R(\boldsymbol{l}_i, \boldsymbol{l}_j; \boldsymbol{\theta})\right)\right) + n_{des}\log(2\pi\sigma^2) + n_{des}\right], \tag{40}$$

where $n_{des}$ is the number of design-of-experiment (DoE) points. Thus, for a number of DoE (training) points, $S_{DoE} = [\boldsymbol{l}_1, \boldsymbol{l}_2, ..., \boldsymbol{l}_m]$, and the corresponding responses from the performance function $\mathbf{Y} = [\text{SOI}(\boldsymbol{l}_1), \text{SOI}(\boldsymbol{l}_2), ..., \text{SOI}(\boldsymbol{l}_m)]$, the traditional BLUP (Best Linear Unbiased Predictor) estimation of Kriging prediction for a group of testing points, $S_t = [\boldsymbol{l}_1, \boldsymbol{l}_2, ..., \boldsymbol{l}_{N_t}]$, gives:

$$\mu_{\hat{k}}(\boldsymbol{l}_t) = \boldsymbol{f}^T(\boldsymbol{l}_t)\widetilde{\boldsymbol{\beta}} + \tilde{r}(\boldsymbol{l}_t)^T \boldsymbol{\gamma}, \qquad \boldsymbol{l}_t \in S_t. \tag{41}$$

where $\boldsymbol{l}_t$ denotes testing samples. Moreover, let $\boldsymbol{C} = \sigma^2 \boldsymbol{R} + \boldsymbol{\Sigma}_n$, $\boldsymbol{\Sigma}_n = \sigma_n^2 \boldsymbol{I}$ (where $\boldsymbol{I}$ is an identity matrix and $\sigma_n^2$ is the variance of noise of SOI) and $\tau = \sigma^2/(\sigma_n^2 + \sigma^2)$, where $\sigma^2$ is the Gaussian process variance, and $\boldsymbol{u}(\boldsymbol{l}_t)$ are:

$$\sigma^2 = \frac{1}{m}(\mathbf{Y} - \boldsymbol{F\beta})^T \boldsymbol{R}^{-1}(\mathbf{Y} - \boldsymbol{F\beta}) \tag{42}$$

The parameters presented in Eq. (41) can be calculated as,

$$\widetilde{\boldsymbol{\beta}} = (\boldsymbol{F}^T \boldsymbol{C}^{-1} \boldsymbol{F})^{-1} \boldsymbol{F}^T \boldsymbol{C}^{-1} \mathbf{Y},$$

$$\boldsymbol{\gamma} = \boldsymbol{C}^{-1}(\mathbf{Y} - \boldsymbol{F}\widetilde{\boldsymbol{\beta}}),$$

$$\boldsymbol{r}(\boldsymbol{l}_t) = [R(\boldsymbol{l}_1, \boldsymbol{l}_t; \boldsymbol{\theta}), .... R(\boldsymbol{l}_m, \boldsymbol{l}_t; \boldsymbol{\theta})]_{1 \times m}^T, \boldsymbol{l}_t \in S_t$$

$$\tilde{\boldsymbol{r}}(\boldsymbol{l}_t) = (1 - \tau)\boldsymbol{r}(\boldsymbol{l}_t)$$

$$\widetilde{R} = (1-\tau)R + \tau I$$

$$F = [f(l_1), f(l_2), \ldots f(l_m)]^T. \tag{43}$$

Then, the mean-square error (MSE) of $\widehat{SOI}(l_t)$ can be calculated by:

$$\sigma_{\hat{k}}^2(l_t) = (\sigma_n^2 + \sigma^2)\left(1 + u^T(l_t)\left(F^T\widetilde{R}^{-1}F\right)^{-1}u(l_t) - \tilde{r}^T(l_t)\widetilde{R}^{-1}\tilde{r}(l_t)\right), \tag{44}$$

where $u(l_t) = F^T C^{-1}r(l_t) - f(l_t)$. According to Kriging theory, for all testing points, $S_t$, the outputs $Y = [g(l_1), g(l_2), \ldots, g(l_t)]$ from the Kriging model are parameterized with the mean, $\mu_{\hat{g}}(l_t)$, and the variance, $\sigma_{\hat{g}}^2(l_t)$:

$$\widehat{SOI}(l_t) \sim N\left(\mu_{\hat{k}}(l_t), \sigma_{\hat{k}}^2(l_t)\right), \quad l_t \in S_t. \tag{45}$$

The general principle of surrogate-based optimization is to start with a small number of training points that compute SOI to build a surrogate for $\widehat{SOI}(L = l)$ and subsequently refine the Kriging surrogate model by adaptively adding new training samples until the target $l^*$ is steadily identified. The procedure discussed above is elaborated in the following steps:

- **Step 1**: Discretizing the regions of observation, $\Omega_{ob}$, into discretized points and denote these samples as $S_{ob}$.

- **Step 2**: Select a limited number of points from $S_\Gamma$ as initial training points $l_{in}$ for Kriging construction. Note that $l_{tr}$ can change upon every iteration of active learning but it is equal to $l_{in}$ in the first iteration.

- **Step 3**: Construct the Kriging model with current training points $l_{tr}$. Denote the Kriging model as $\widehat{SOI}(l)$. Construction is based on UQLab toolbox in MATLAB®, with ordinary Kriging basis and Gaussian correlation function. The model type is selected as prediction with noisy responses and other parameters follow default settings. Subsequently, the Kriging responses $\mu_{\hat{k}}(l)$ and variances $\sigma_{\hat{k}}^2(l)$ can be acquired from UQLab toolbox ("UQLab Kriging (Gaussian process modelling) manual," n.d.).

- **Step 4**: To search for the maximum value of SOI, the expected improvement learning function (EI) for global optimization is adopted. The next training point is selected according to the following criterion and is denoted as $l_{tr}^*$.

$$l_{tr}^* = \arg\max_{l \in S_\Gamma} EI(l) \tag{46}$$

where,

$$EI(l) = \left(\mu_{\hat{k}}(l) - \widehat{SOI}_{max}^*\right)\Phi\left(\frac{\mu_{\hat{k}}(l) - \widehat{SOI}_{max}^*}{\sigma_{\hat{k}}(l)}\right) + \sigma_{\hat{k}}(l)\varphi\left(\frac{\mu_{\hat{k}}(l) - \widehat{SOI}_{max}^*}{\sigma_{\hat{k}}(l)}\right) \tag{47}$$

where $\widehat{SOI}_{max}^*$ denotes the maximum SOI among $l_{tr}$ in the current iteration.

- **Step 5**: Determine if the stopping criterion ($max(EI) \leq EI_{thr}$) has been satisfied in the current iteration, where $EI_{thr}$ denotes the threshold value. In this paper, $EI_{thr}$ is set as $10^{-5}$ and the maximum number of iterations is set as 100. Go to Step 6 if satisfied; otherwise, go back to Step 3.

- **Step 6:** Output $l^*$ and $\widehat{SOI}$ for $S_{ob}$.

The above procedure presents an efficient approach for evaluating SOI for all potential monitoring points. However, the computational complexity of constructing Kriging surrogate model increases substantially as $N_{dim}$ grows. This is known as the 'curse of dimensionality', which can be further optimized in the future. To explore the performance of the proposed framework, a geotechnical case that investigates tunneling-induced settlement to building damage is investigated in the next section.

## 4 Case study
### 4.1 Description of the physical model
Settlements caused by the construction of tunnel can threaten the functionalities and integrity of structures and infrastructure systems overground. This process is illustrated in Fig 2, where the *y* axis follows the reverse direction of tunnel advance and *x* axis is perpendicular to the tunnel longitudinal axis (Camós et al., 2016; Camós and Molins, 2015). Origin is the intersection of the extensional line of building wall and the *y*-axis. Moreover, the positive degree refers to alignments counterclockwise with respect to the *x*-axis. Starting from tunnel portal $y = y_f$ and adaptively advancing towards $y = -\infty$ with the tunnel boring machine (TBM), a tunnel is under construction with the tunnel face located at $y = y_s$. In this paper, $y_f$ is assumed to originate from infinity with $y_f = +\infty$. A building wall of length $l_{build}$, denoted by a reference point $\hat{A}$, is located at a distance $d_{orig}$ from the origin and aligned $\theta_r$ degrees with respect to the tunnel transverse plane. To better interpret the concept, a 3D plot of this physical model is showcased in Fig 3, where $d$ and $z_0$ denote the diameter of tunnel and the depth from surface of ground ($z = 0$) to the center of tunnel.

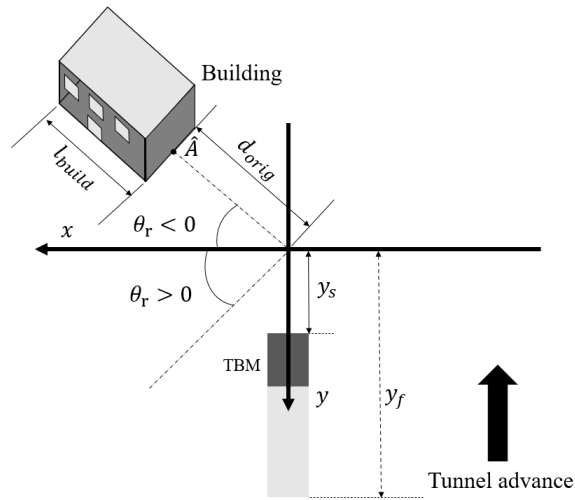

**Fig 2.** Illustration of tunneling-induced settlements.

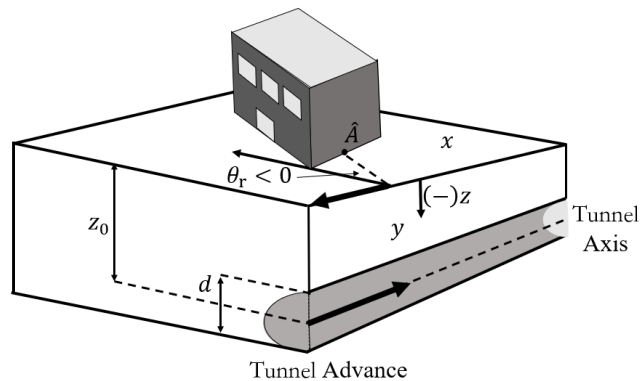

**Fig 3.** 3D illustration of tunnel and building wall positions.

The soil over the excavated underground space can be viewed as a distributed loading with the other ending node fixed at $y = +\infty$. According to (Attewell and Woodman, 1982; Camós et al., 2016; O'reilly and New, 1982), the settlement of ground can be calculated as,

$$S(x, y, z, d, y_s, y_0, y_f, z_0, V_L, K_x, K_y)$$
$$= -1000 \cdot S_{max} \cdot \exp\left[-\frac{x^2}{2 \cdot K_x^2 \cdot (z_0 - z)^2}\right]$$
$$\cdot \left[\Phi\left(\frac{y - (y_s + y_0)}{K_y \cdot (z_0 - z)}\right) - \Phi\left(\frac{y - y_f}{K_y \cdot (z_0 - z)}\right)\right] \quad (48)$$

where $V_L$ is the volume ground loss per unit and $K_x$ and $K_y$ are non-dimensional trough width parameters reflecting characteristics of the soil and describing the Gaussian settlement profiles in the transverse and longitudinal direction. It is typically assumed that $K_x = K_y = K$ (Attewell et al., 1986). Moreover, $S_{max}$ denotes the absolute value of maximum settlement at $y$ ($y \geq y_s$) and can be calculated as,

$$S_{max} = \frac{V_L \cdot \pi \cdot d^2}{\sqrt{2\pi} \cdot K_x \cdot (z_0 - z) \cdot 4} \quad (49)$$

$y_0$ in Eq. (48) is the horizontal shift of the longitudinal settlement profile with respect to the tunnel face and can be calculated as,

$$y_0 = -\Phi^{-1}(\delta) \cdot K \cdot z_0 \quad (50)$$

where $\delta$ represents the ratio between the surface settlement above the tunnel face and $S_{max}$ at $y = +\infty$. In this paper, $\delta$ is defined as 0.3 for the sake of practical consideration [40], [41]. The shape of settlement is presented in Fig 4. It can be observed that the shape of settlement along the x-axis follows the PDF of Gaussian distribution, while along the y-axis, the shape is close to the CDF of Gaussian distribution. The settlement reaches the highest value at (0,30,0).

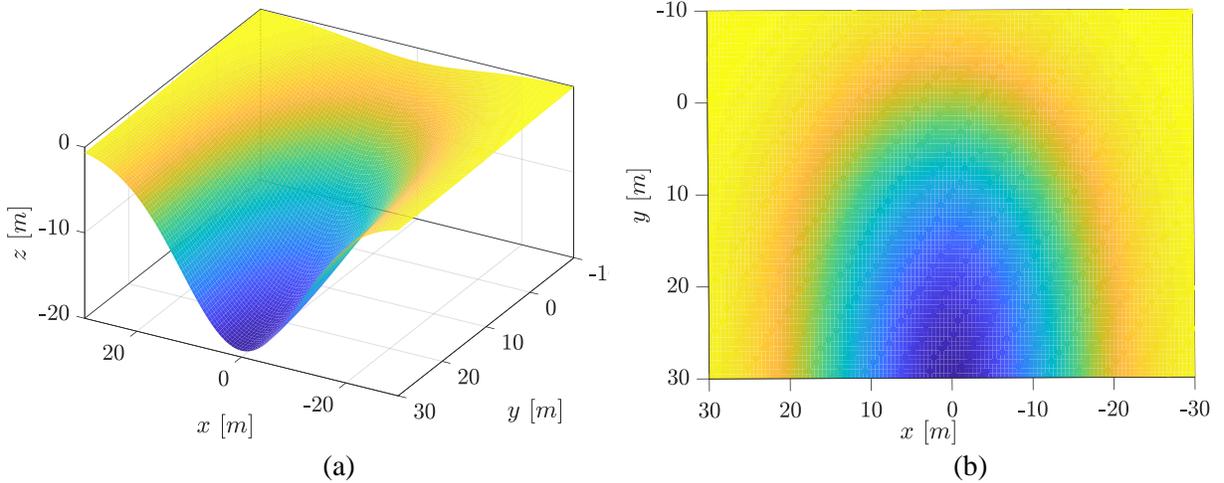

**Fig 4.** Settlement produced by tunnel excavation in (a) the 3D surface plot and (b) the x-y view with $d = 12$m, $y_s = 0$, $z_0 = 23$m, $V_L = 0.5\%$ and $K = 0.5$.

By treating the building wall as a weightless linear elastic rectangular beam, the response of the building to the settlement is modeled through the equivalent beam method (Camós et al., 2014). The distribution of tensile strains along the beam is dominated by the shape of the deflection and the mode of deformation. The extreme fiber strains caused by bending and shear $\varepsilon_{br}$ and $\varepsilon_{dr}$ can be calculated according to the following equations,

$$\varepsilon_{br}\left(V_L, K, \frac{E}{G}\right) = (\varepsilon_{b\ max} + \varepsilon_h) \cdot E_{\varepsilon_{br}} \tag{51}$$

$$\varepsilon_{dr}\left(V_L, K, \frac{E}{G}\right) = \left[\varepsilon_h\left(1 - \frac{E}{4G}\right) + \sqrt{\frac{\varepsilon_h^2}{16}\left(\frac{E}{G}\right)^2 + \varepsilon_{d\ max}^2}\right] \cdot E_{\varepsilon_{dr}} \tag{52}$$

where $\frac{E}{G}$ represents the ratio between the Young's modulus and the shear modulus of the building material and $E_{\varepsilon_{br}}$ and $E_{\varepsilon_{dr}}$ are multiplicative model errors. In this paper, $\frac{E}{G}$, $E_{\varepsilon_{br}}$ and $E_{\varepsilon_{dr}}$ are modeled as random variables. Moreover, $\varepsilon_{b\ max}$ and $\varepsilon_{d\ max}$ are the maximum bending and shear strains due to the deflection. Specifically, $\varepsilon_{b\ max}$ and $\varepsilon_{d\ max}$ are calculated separately for the different zones of the building. The building zones that have settlements induced by the tunnel can be typically classified into two types: the sagging and hogging deflections. The main difference of them lies in the position of the profile curvature change: sagging deflection represents upwards concavity while hogging deflection indicates downwards concavity. To better illustrate the foregoing difference, Fig 6 showcases the sagging and hogging deflections in the different zones of a building.

The number of inflection points along the building depends on the three parameters $l_{build}$, $d_{orig}$ and $\theta_r$. Moreover, the type of deflection of a building depends on the number of inflection points, which are summarized in Table 1. A conceptual plot for the last case in Table 1 is shown in Fig 6, where the building is divided in to three zone: one sagging zone and two hogging zones. Let $l_{ref}$ denote the horizontal distance between two reference points and $\Delta_{ref}$ be the relative deflection; the deflection ration $\Delta_{ref}/l_{ref}$ for different deflection types can be represented as $\Delta_{sag}/l_{sag}$ and $\Delta_{hog}/l_{hog}$. The maximum bending and shear strains, $\varepsilon_{b\ max}$ and $\varepsilon_{d\ max}$, for a given zone (sagging or hogging) can be calculated as follows (Burland and Wroth, 1975),

$$\varepsilon_{b\ max} = \frac{\Delta_{ref}/l_{ref}}{\left(\frac{l_{ref}}{12t} + \frac{3I}{2al_{ref}H}\frac{E}{G}\right)} \tag{53}$$

$$\varepsilon_{d\ max} = \frac{\Delta_{ref}/l_{ref}}{\left(1 + \frac{Hl_{ref}^2}{18I}\frac{G}{E}\right)} \tag{54}$$

where $H$ denotes the height of building, $I = H^3/12$ is the inertia per unit length, $t$ is depth of neutral axis and $a = t$ is the location of the fiber where strains are calculated. For sagging and hogging deflections, $t = H/2$ and $H$, respectively. Moreover, the resultant horizontal strain in the ground surface along the base of the team, $\varepsilon_h$, in Eq. (51) and (52) can be computed according to the following equation:

$$\varepsilon_h(x, y, z, V_L, K) \equiv \cos^2\theta_r \cdot \varepsilon_{h,xx} + \sin^2\theta_r \cdot \varepsilon_{h,yy} + 2 \cdot \cos\theta_r \sin\theta_r \cdot \varepsilon_{h,xy} \tag{55}$$

where $\varepsilon_{h,xx}$, $\varepsilon_{h,yy}$ and $\varepsilon_{h,xy}$ are the fields of strain in the ground, . The maximum strain of the building $\varepsilon_{max}$ can be determine according to the six parameters,

$$\varepsilon_{max} = \max[\varepsilon_{br}^{sag}, \varepsilon_{dr}^{sag}, \varepsilon_{br}^{hog,1}, \varepsilon_{dr}^{hog,1}, \varepsilon_{br}^{hog,2}, \varepsilon_{dr}^{hog,2}] \qquad (56)$$

**Table 2.** Classification of damage (Burland and Wroth, 1975).

| Category of damage | Normal degree of severity | Typical damage | Tensile strain $\varepsilon_{max}$(%) | $\varepsilon_{lim}$(%) |
|---|---|---|---|---|
| 0 | Negligible | < 0.1 mm | 0-0.050 | 0.050 |
| 1 | Very slight | < 1.0 mm | 0.050-0.075 | 0.075 |
| 2 | Slight | < 5.0 mm | 0.075-0.150 | 0.150 |
| 3 | Moderate | < 15.0 mm | 0.150-0.300 | 0.300 |
| 4 | Severe | < 25.0 mm | >0.300 | - |
| 5 | Very Severe | > 25.0 mm | - | - |

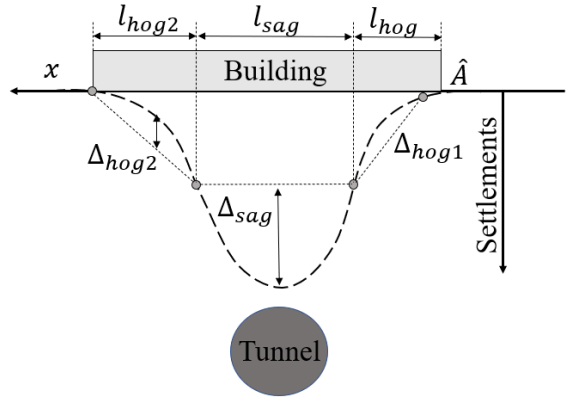

**Fig 5.** Conceptual illustration of sagging and hogging deflections in different zones of a building (Camós et al., 2016).

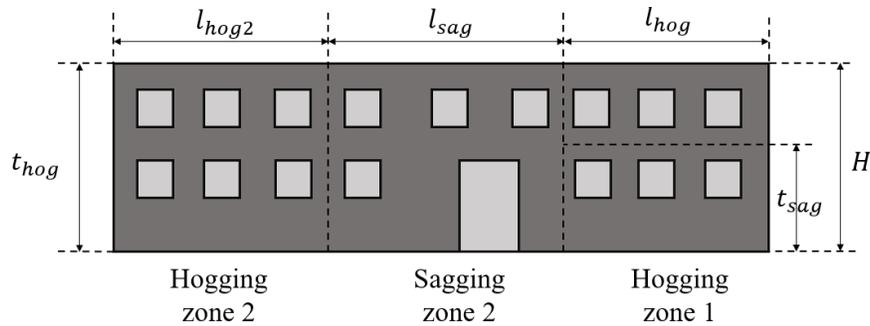

**Fig 6.** A building that is subjected to tunneling-induced settlement with 1 sagging and 2 hogging zones (Camós et al., 2016).

The location of the building determines the number of extreme fiber strains. It indicates that the building can be divided into sagging and hogging zones. Therefore, six notations including $\varepsilon_{br}^{sag}$, $\varepsilon_{dr}^{sag}$, $\varepsilon_{br}^{hog,1}$, $\varepsilon_{dr}^{hog,1}$, $\varepsilon_{br}^{hog,2}$ and $\varepsilon_{dr}^{hog,2}$ are sufficient to represent the 4 cases described in Table 1, where $\varepsilon_{br}^{sag}$, $\varepsilon_{br}^{hog,1}$ and $\varepsilon_{br}^{hog,2}$ are the maximum bending strains in sagging zone and $\varepsilon_{dr}^{sag}$, $\varepsilon_{dr}^{hog,1}$ and $\varepsilon_{dr}^{hog,2}$ are the maximum shear strains in hogging zone. For the first case, the last four terms are equal to zero due to the existence of one sagging zone; For the second case, one hogging zone indicates that the third and fourth terms are non-zero; For the third case, the last two terms are equal to zero while all the six terms are non-zero for the last case. Finally, the geometry of the cracks can be estimated. Moreover, the classification of damage due the cracks is summarized in Table 2, where $\varepsilon_{lim}$ denotes the limit tensile strain.

**Table 1.** Types of deflection of building located in a specific location

| Location of building | Number of inflections | Types of Deflection |
|---|---|---|
| Above the tunnel axis | 0 | 1 sagging |
| Far from the tunnel axis | 0 | 1 hogging |
| Starts in the sagging zone and reaches the hogging zone | 1 | 1 sagging and 1 hogging |
| Central part in the sagging zone and lateral parts in the hogging zone | 2 | 1 sagging and 2 hogging |

To calculate $\varepsilon_{h,xx}$, $\varepsilon_{h,yy}$ and $\varepsilon_{h,xy}$, let $U_x$ and $U_y$ denote the horizontal displacements in [mm] in the transvers and longitudinal direction, respectively, at a certain position with coordinate $x, y, z$ in [m]. $U_x$ and $U_y$ can be calculated as,

$$U_x = \frac{x}{z_0 - z} \cdot S \tag{57}$$

and

$$U_y = 1000 \cdot \frac{V_L \cdot d^2}{8 \cdot (z_0 - z)} \cdot \left[ \exp\left( \frac{-(y - (y_s + y_0))^2 - x^2}{2 \cdot K_y^2 \cdot (z_0 - z)^2} \right) - \exp\left( \frac{-(y - y_f)^2 - x^2}{2 \cdot K_y^2 \cdot (z_0 - z)^2} \right) \right] \tag{58}$$

Therefore $\varepsilon_{h,xx}$, $\varepsilon_{h,yy}$ and $\varepsilon_{h,xy}$ can be calculated based on $U_x$ and $U_y$,

$$\varepsilon_{h,xx} = \frac{\partial U_x}{\partial x} = \frac{S/1000}{z_0 - z} \cdot \left( 1 - \left( \frac{x^2}{K_x^2 \cdot (z_0 - z)^2} \right) \right) \tag{59}$$

and

$$\varepsilon_{h,yy} = \frac{\partial U_y}{\partial y} = \frac{V_L \cdot d^2}{8 \cdot (z_0 - z)} \cdot \left[ \begin{array}{l} \left(\frac{-2y + 2(y_s + y_0)}{2 \cdot K_y^2 \cdot (z_0 - z)^2}\right) \exp\left(\frac{-(y - (y_s + y_0))^2 - x^2}{2 \cdot K_y^2 \cdot (z_0 - z)^2}\right) \\ - \left(\frac{-2y + 2y_f}{2 \cdot K_y^2 \cdot (z_0 - z)^2}\right) \exp\left(\frac{-(y - y_f)^2 - x^2}{2 \cdot K_y^2 \cdot (z_0 - z)^2}\right) \end{array} \right] \quad (60)$$

and

$$\varepsilon_{h,xy} = \frac{1}{2}\left(\frac{\partial U_x}{\partial y} + \frac{\partial U_y}{\partial x}\right) \quad (61)$$

where $\frac{\partial U_x}{\partial y}$ and $\frac{\partial U_y}{\partial x}$ read as,

$$\frac{\partial U_x}{\partial y} = \frac{x}{z_0 - z} \cdot \left(-\frac{V_L \cdot \pi \cdot d^2}{\sqrt{2\pi} \cdot K_x \cdot (z_0 - z) \cdot 4}\right) \cdot \left( \begin{array}{l} \frac{1}{\sqrt{2\pi}} e^{-\frac{\left(\frac{y-(y_s+y_0)}{K_y(z_0-z)}\right)^2}{2}} \cdot \left(\frac{1}{K_y(z_0-z)}\right) \\ -\frac{1}{\sqrt{2\pi}} e^{-\frac{\left(\frac{y-y_f}{K_y(z_0-z)}\right)^2}{2}} \cdot \left(\frac{1}{K_y(z_0-z)}\right) \cdot \exp\left(-\frac{x^2}{2 \cdot K_x^2 \cdot (z_0-z)^2}\right) \end{array} \right) \quad (62)$$

and

$$\frac{\partial U_y}{\partial x} = \frac{V_L \cdot d^2}{8 \cdot (z_0 - z)} \cdot \frac{(-2x)}{2 \cdot K_x^2 \cdot (z_0 - z)^2} \left[ \begin{array}{l} \exp\left(\frac{-(y - (y_s + y_0))^2 - x^2}{2 \cdot K_y^2 \cdot (z_0 - z)^2}\right) \\ -\exp\left(\frac{-(y - y_f)^2 - x^2}{2 \cdot K_y^2 \cdot (z_0 - z)^2}\right) \end{array} \right] \quad (63)$$

### 4.2 Sensitivity of information (SOI) analysis

In this subsection, the computational procedure of the estimate of SOI is elaborated at different locations. First, the tunnel excavation-caused building crack larger than 0.1mm ($\varepsilon_{lim} = 0.05\%$) is defined as structurally intolerable, indicating the damage level 1 in Table 2. Accordingly, the limiting strain for this case can be set as $\varepsilon_{lim} = 0.05\%$, leading to the limit state function (LSF), $g(X)$ for this case,

$$g(X) = \varepsilon_{lim} - \varepsilon_{max}(X) \quad (64)$$

where $X$ denotes the vector of random variables. In this context, $X = \left[V_L; K; \frac{E}{G}; E_{\varepsilon_{br}}^{sag}; E_{\varepsilon_{br}}^{hog,1}; E_{\varepsilon_{br}}^{hog,2}; E_{\varepsilon_{dr}}^{sag}; E_{\varepsilon_{dr}}^{hog,1}; E_{\varepsilon_{dr}}^{hog,2}\right]$, where $E_{\varepsilon_{br}}^{sag}, E_{\varepsilon_{br}}^{hog,1}, E_{\varepsilon_{br}}^{hog,2}, E_{\varepsilon_{dr}}^{sag}, E_{\varepsilon_{dr}}^{hog,1}$ and $E_{\varepsilon_{dr}}^{hog,2}$ are the errors of the equivalent beam model of Eq. (51) and (52) in the sagging and hogging zones, respectively. The probabilistic distribution of these 9 random variables is summarized in Table 3. Moreover, the failure domain $\Omega_f$ can be defined as,

$$\Omega_f = \{g(x) \leq 0\} \quad (65)$$

where $x$ is a stochastic realization from $X$. To precisely assess and track the risk of the tunnelling-induced settlement to the building, the measurement of settlement at the location $l_m$, $s_m(l_m)$, is conducted over the region of observation, $\Omega_{ob}$. $Z_{lob}$ and $Z_{upb}$ are equal to 5 and 15, respectively, based on engineering experience. Fig 7 illustratively interpret this strategy, where the light green region is represented as $\Omega_{ob}$. Moreover, the relation between the measured and ground truth settlement can be read

as:

$$S_m = S(x_m^i, y_m^i, z_m^i, V_l, K) + E_f + E_m = S(x_m^i, y_m^i, z_m^i, V_l, K) + E_E \tag{66}$$

where $E_f$ is the model error interpreting the potential inaccuracy of the Gaussian settlement shape and $E_m$ is the measurement error stemming from the manmade imprecision, imperfection of instruments etc. Let $E_E = E_f + E_m$, the likelihood function for this case is defined as:

$$L(v_l, k) = f_E \left( s_m^i - S(x_m^i, y_m^i, z_m^i, v_l, k) \right) \tag{67}$$

where $v_l$ and $k$ are the realizations of random variables $V_l$ and $K$, and $f_E$ is the PDF of the integrated error $E_E$. Therefore, the two augmented limit state function $h_1(P, X)$ and $h_2(P, X)$ can be sequentially defined based on Eq. (67). Before exploring $l^*$ over $\Omega_{ob}$, the procedures of estimating P(F) and P(F|Z) are elaborated. Therein, four measurements, $s_m(x_m^i, y_m^i, z_m^i), i = 1,2,3$ and 4 are ready to update P(F|Z) from P(F). The corresponding data and simulation results are reported in Table 4 together with the corresponding interpretative figures illustrated in Fig 8.

**Table 3.** Probabilistic distribution of random variables (Camós et al., 2016).

| Random variable | Description | Type of distribution | Mean | Standard deviation |
|---|---|---|---|---|
| $K(-)$ | Trough width parameter | Lognormal | 0.3 | 0.06 |
| $V_L(\%)$ | Volume loss | Lognormal | 0.4 | 0.16 |
| $\frac{E}{G}(-)$ | Material ratio | Beta | 2.5 | 0.045 |
| $\begin{bmatrix} E_{\varepsilon_{br}}^{sag}; E_{\varepsilon_{br}}^{hog,1}; E_{\varepsilon_{br}}^{hog,2}; \\ E_{\varepsilon_{dr}}^{sag}; E_{\varepsilon_{dr}}^{hog,1}; E_{\varepsilon_{dr}}^{hog,2} \end{bmatrix}(-)$ | Equivalent beam model errors | Lognormal | 1.0 | 0.05 |
| $E_m(mm)$ | Measurement error | Normal | 0 | 1 |
| $E_f(mm)$ | Settlement model error | Normal | 0 | 2 |

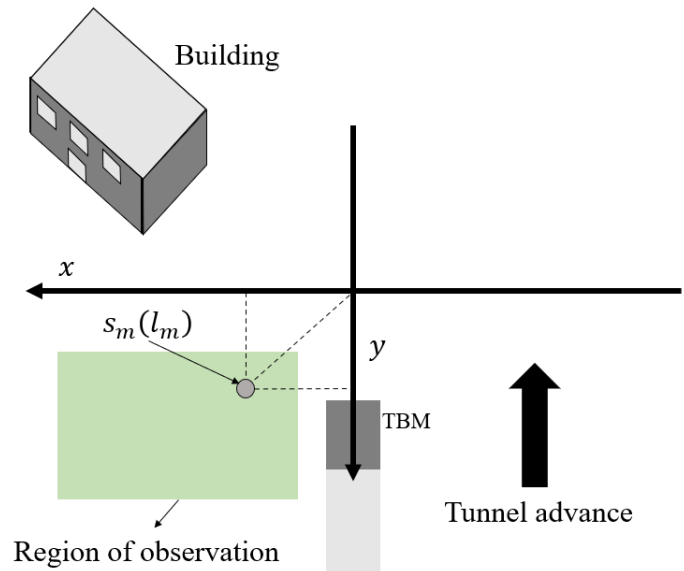

**Fig 7.** Conceptual illustration of $\Omega_{ob}$ and the monitoring measurement $s_m(l_m)$, where $l_m = (x_m, y_m, z_m)$ and $l_m \in \Omega_{ob}$.

**Table 4.** Simulation results of case study through the proposed reliability updating method, where $COV_{thr} = 0.05$ and $N_{in} = 10^4$.

| Information | Position | Pr(F) | Pr(F\|Z) | $r_{up}$ | $COV_{P_F}$ | $COV_{P_{F\|Z}}$ | $N_{eva}$ |
|---|---|---|---|---|---|---|---|
| $s_m(l_1) = 10$ | (10,10,0) | $8.26 \times 10^{-3}$ | $1.29 \times 10^{-2}$ | 0.069 | 0.0449 | 0.0471 | 56000 |
| $s_m(l_2) = 10$ | (15,15,0) | $8.40 \times 10^{-3}$ | $8.47 \times 10^{-2}$ | 0.425 | 0.0375 | 0.0497 | 84000 |
| $s_m(l_3) = 10$ | (20,20,0) | $8.31 \times 10^{-3}$ | $2.23 \times 10^{-2}$ | 0.161 | 0.0448 | 0.0494 | 56000 |
| $s_m(l_4) = 10$ | (25,25,0) | $8.36 \times 10^{-3}$ | $9.84 \times 10^{-3}$ | 0.025 | 0.0453 | 0.0462 | 56000 |

By setting $K$ and $V_L$ as the $x$ and $y$ axis and starting with $N_{ss} = 10^4$, Fig 8(a) illustrates the estimate of Pr(F) through $SS$ with information $s_m(l_4)$, where $S_1$, $S_2$ and $S_3$ denote samples located in the three intermediate subsets. However, the initial set of $N_{ss}$ is insufficiently large so that $COV_{P_F}$ is estimated as large as 0.0657. Therefore, $N_{ss}$ is adaptively increased to $2 \times 10^4$ and Pr(F) is finally estimated as $8.26 \times 10^{-3}$ with $COV_{P_F}$ equal to 0.0449. Fig 8(b) showcases the estimate of Pr(Z) based on the augmented limit state function $h_1(P, X) = 0$, where the darker dots denote the accepted samples, $S_{acc}$, and the brighter ones represent the rejected samples, $S_{rej}$. In this step, Pr(Z) is estimated as $8.87 \times 10^{-6}$ and $c_1$ is equal to $1.78 \times 10^{-5}$. Fig 8(c) showcases the estimate of Pr(Z|F) through the augmented limit state function $h_2(P, X) = 0$, where the darker dots denote the accepted samples, $S_{acc}^{last}$, and the brighter ones represent the rejected samples, $S_{rej}^{last}$. One should note that $[S_{acc}^{last}, S_{rej}^{last}] \in S^{last}$, where $S^{last}$ is the last sample set in Fig 8(a). In this step, the two terms are estimated as $Pr(Z|F) = 1.05 \times 10^{-5}$ and $c_2 = 2.02 \times 10^{-5}$. Fig 8(d) exhibits the conventional procedure represented in Eq. (12) that relies on the computational scheme $P(F \cap Z)/P(Z)$ with the joint limit state function $J(P, X) = \max[g(X), h_1(P, X)]$. The conventional approach results in the simulation data with $P(F \cap Z) = 2.92 \times 10^{-3}$, which is significantly smaller than Pr(F). This implies that more evaluations of $g(X)$ should be conducted compared to the proposed approach, which further demonstrates the computational efficiency of the proposed reliability updating approach. Moreover, the size of samples in each subset is adaptively increased to guarantee the sufficiently consistency of Pr(Z|F), which facilitates the computational robustness of SOI and the exploration of $l^*$. Fig 9 showcases the relation of $r_{up}$ versus Z at locations $l_1$ and $l_3$. One can infer that $r_{up}$ reaches 0 at location $l_1$ when $Z = 8.7$, which indicates that the updated reliability deviates substantially from the prior one when the information is involved because it is beyond the expectation of prior knowledge. However, $r_{up}$ almost increases linearly at location $l_3$ over the interval $[Z_{lob}, Z_{upb}]$. The next subsection elaborates the procedures of exploring $l^*$ over different regions of observation and excavation stage.

**Table 5.** SOIs at location $l_1, l_2, l_3$ and $l_4$

| Number | Location | SOI |
|---|---|---|
| 1 | $l_1$ | 0.0259 |
| 2 | $l_2$ | 0.0403 |
| 3 | $l_3$ | 0.0157 |
| 4 | $l_4$ | 0.0027 |

It can be observed from Table 4 that $r_{up}$ for the information $s_m(l_2) = 10$ is apparently the largest one, which also indicates the largest change for the update of reliability when information $s_m(l_2)$ is available. As the location transits from $l_2$ to $l_4$ and the settlement information keeps unchanged, $r_{up}$ changes significantly from 0.425 to 0.025. This is in attributed to the fact that the location of $l_4$ is further from both the tunnel axis and building facade compared to location $l_2$. Therefore, $r_{up}$ can be an efficient metric for quantifying the contribution of the change of updated reliability for different source of information. Moreover, the significance of information at some locations cannot be interpreted by intuition, therefore, the metric $r_{up}$ can be utilized to facilitate this process. For example, $r_{up}$ is estimated as 0.069 at location $l_1$, which is less significant than location $l_3$ because the settlement close to TBM becomes smaller, thereby it has less influence on $r_{up}$. This point, however, does not indicate that $l_1$ is less valuable than $l_3$ in terms of SOI. For example, Table 5 presents that SOI for location $l_1$ is greater than location $l_3$. Nevertheless, $l_2$ is deemed to be the most significant location as represented by the largest SOI.

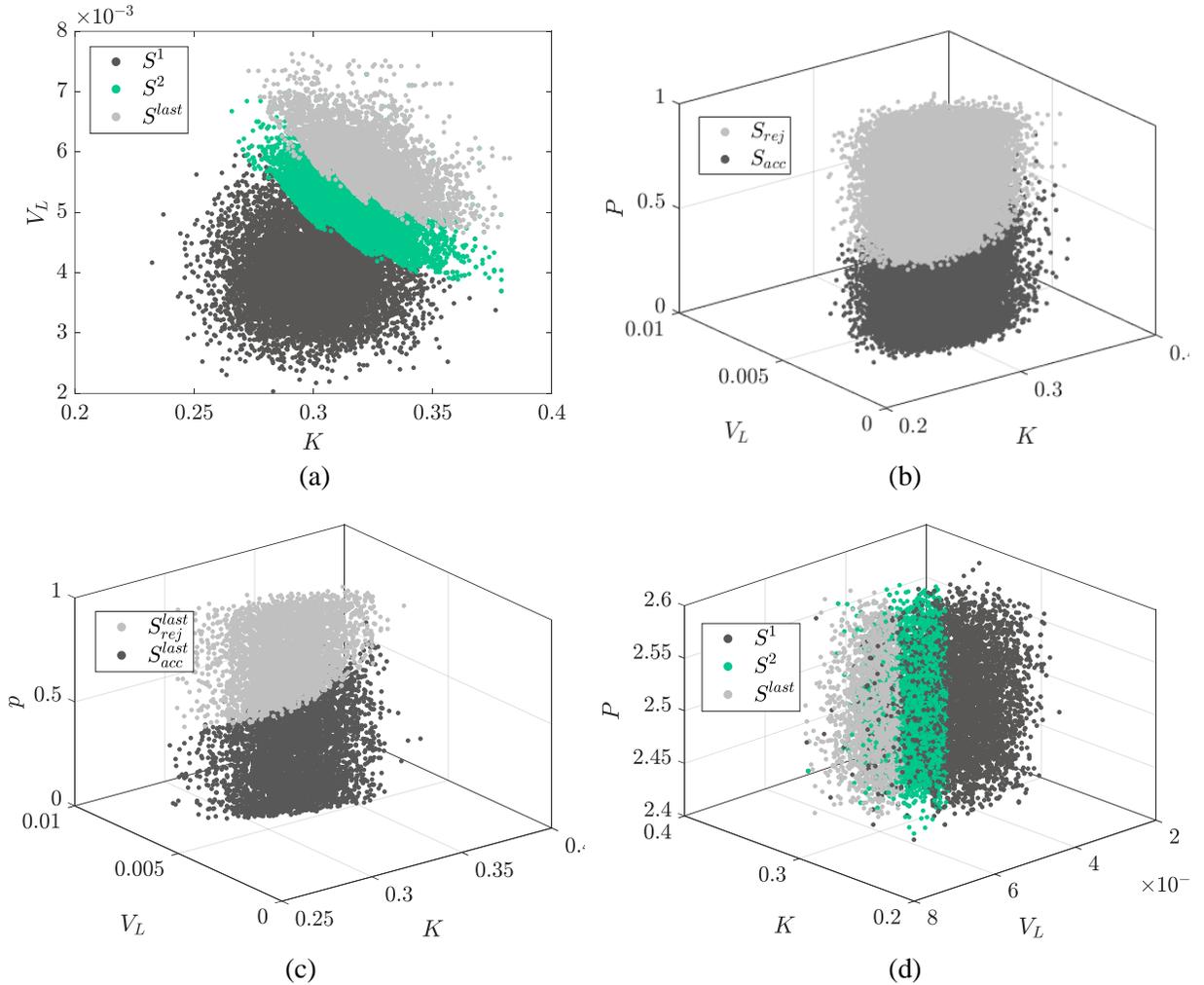

**Fig 8.** Use information $s_m(l_4)$ via *SS* to estimate (a) Pr(F) based on $g(\boldsymbol{X})$; (b) Pr(Z) based on $h_1(P, \boldsymbol{X})$ at the location; (c) Pr(Z|F) based on $h_2(P, \boldsymbol{X}')$, where $\boldsymbol{X}' \in \Omega_f$ and (d) Pr(Z ∩ F) based on $J(P, \boldsymbol{X}) = \max[g(\boldsymbol{X}), h_1(P, \boldsymbol{X})]$.

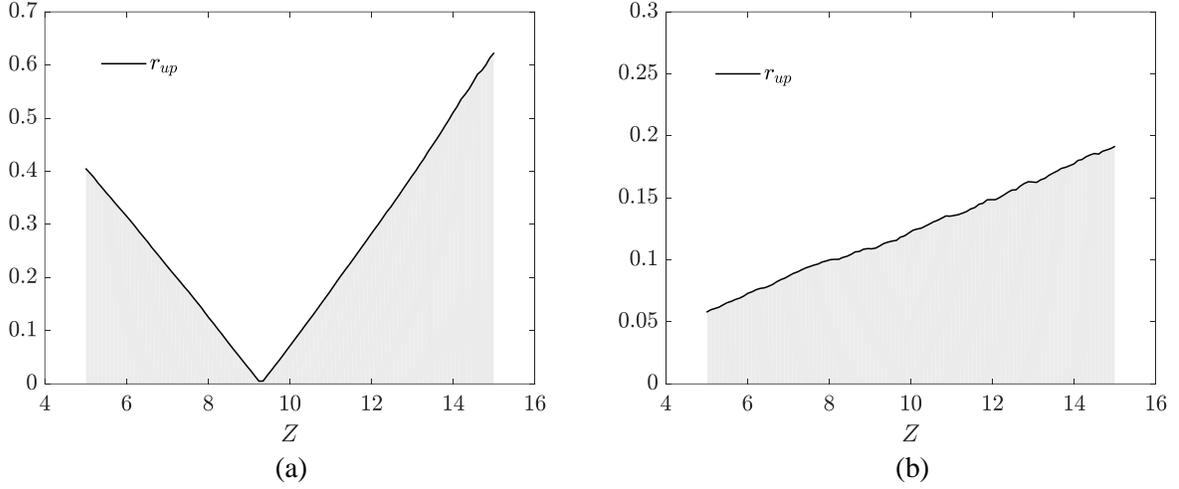

**Fig 9.** $r_{up}$ versus Z at the location of (a) $l_1$ and (b) $l_3$

### 4.3 Optimal settlement monitoring location

In this subsection, $l^*$ is explored over the whole region of observations, $\Omega_{ob}$, globally, where the corresponding steps are depicted in Fig10. In Fig 10(a), the tunnel is plotted with gray region, the light red square showcases where the building façade locates, $C_s$ represents the contour of tunneling excavation-induced settlement and $\Omega_{ob}$ is represented by light blue square. Initially, 81 equally distributed training samples (locations), $l_{in}$, represented with black star dots are ready to training a surrogate model for $\widehat{SOI}(l)$ over $\Omega_{ob} = [x^1_{lim}, x^2_{lim}; y^1_{lim}, y^2_{lim};]$, where $[x^1_{lim}, x^2_{lim}; y^1_{lim}, y^2_{lim};]$ denotes the x and y limits of axis of $\Omega_{ob}$. For example, $\Omega_{ob}$ is parameterized by $[10,30; 10,30;]$ in the Fig 10(a) and the true responses of the 81 discretized training samples are estimated, which facilitates the initial construction of $\widehat{SOI}(l)$. Subsequently, extra training samples are adaptively enriched through the EI active learning function and terminates until the stopping criterion is satisfied. For this case, 15 extra training samples are finally added and the surface plot of $\widehat{SOI}(l)$ based on the Kriging surrogate model is represented in Fig 10(b). Fig 10(c) showcases the x-y view of $\widehat{SOI}(l)$, where $\widehat{SOI}$ increase as the regions transits from blue to yellow. Moreover, the initial optimal location, $l^*_{in}$, among $l_{in}$ is identified as $(15,25,0)$ and finally transits to the final optimal location, $l^*$, where $l^* = (14.7, 24.2, 0)$ with SOI estimated as 0.0513, as highlighted in Fig 10(d).

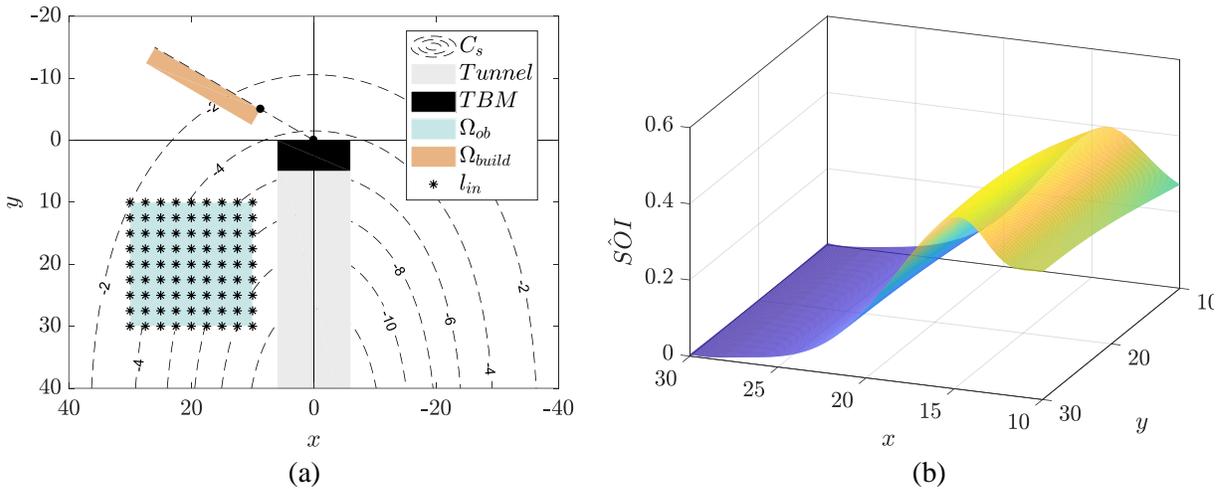

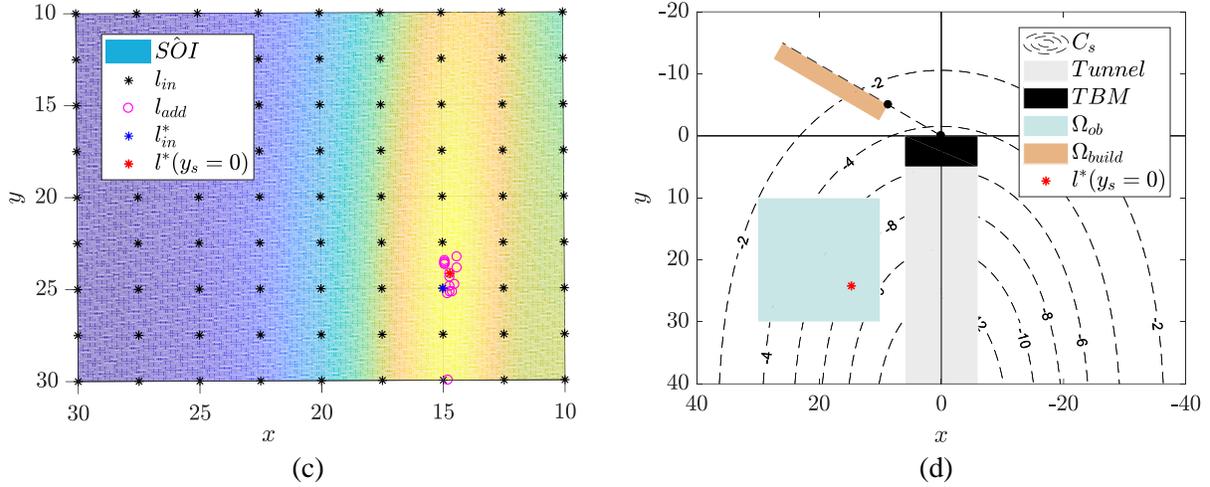

|          |          |
|----------|----------|
| (c)      | (d)      |

**Fig 10.** Procedure of exploring $l^*$ through surrogate-based optimization with (a) the initial training samples located in $\Omega_{ob} = [10,30; 10,30;]$, where $y_s = 0$; (b) Surface plot of $\widehat{SOI}$; (c) the add training samples through active learning and finally identified $l^*$ and (d) an overview of $l^*$ in the process of tunneling excavation.

As the excavation of tunnel proceeds and $\Omega_{ob}$ changes, $l^*$ changes accordingly. Fig 11 showcases four scenarios, $\Omega_A, \Omega_B, \Omega_C$ and $\Omega_d$ of $\Omega_{ob}$ for tunnel excavation along with the tunnel façade $y_s$ changing from 0 to -3, of which the simulation results are summarized in Table 6. According to Fig 11 and Table 6, $l^*$ changes from $(14.7, 24.2, 0)$ to $(12.5, 20.2, 0)$ while the excavation proceeds to $y_s = -3$ and $\Omega_{ob}$ keeps unchanged which leads to SOI changes from 0.0513 to 0.1435. This phenomenon can be interpreted by the fact that the settlement caused by excavation of tunnel dominates the change of the updated reliability. Moreover, Fig 11(b) represents that $l^*$ maintains unchanged even though $y_s$ changes from 0 to -3 along with the increase of SOI from 0.0069 to 0.0163. This is because the point at the very bottom left over $\Omega_{ob} = \Omega_B$ is the most valuable point. As $\Omega_{ob}$ changes from $\Omega_A$ to $\Omega_B$, $l^*(y_s = 0)$ and $l^*(y_s = -3)$ are estimated as $(-14.5, 23.7, 0)$ and $(-11.9, 21.5, 0)$ with SOI equals to 0.0498 and 0.1511, respectively. It can be inferred from the comparison between Fig 11(a) and (c) that above two optimal location are closely symmetric to the two identified optimal locations when $\Omega_{ob} = [10,30; 10,30;]$ along the y-axis. This can be explained by the symmetric characteristics of Gaussian settlement defined in Eq. (48). In Fig 12(d), $l^*$ changes from $(0.5, -10, 0)$ to $(7.5, -10, 0)$ with the corresponding SOI estimated as 0.0133 and 0.0052, when $\Omega_{ob} = \Omega_D$. The tunnel façade at $y_s = 0$ causes a slight deviation of $l^*$ close to the building side when $\Omega_{ob} = \Omega_D$ and this effect of deviation strengthens when $y_s = -3$.

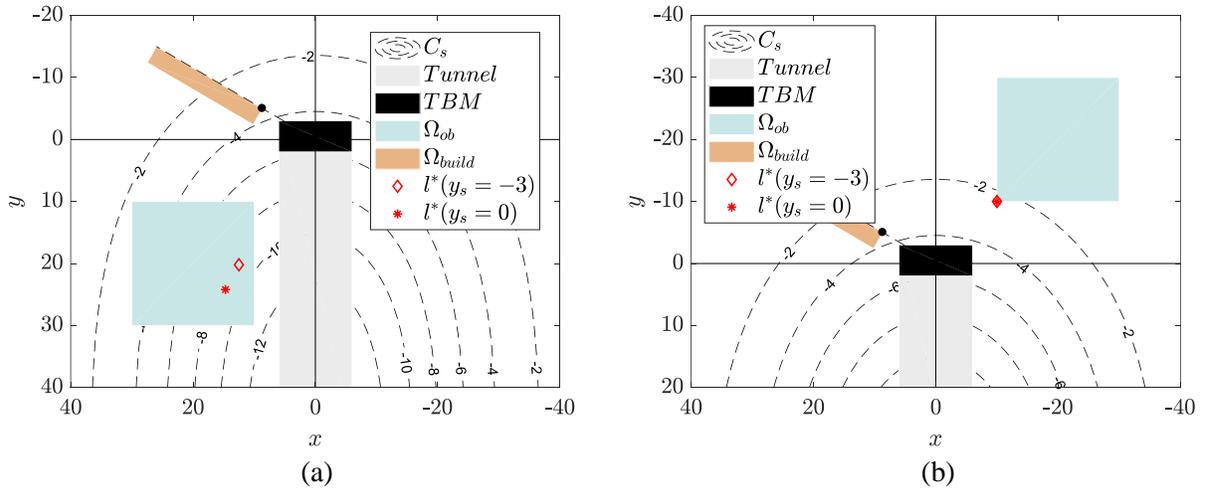

|          |          |
|----------|----------|
| (a)      | (b)      |

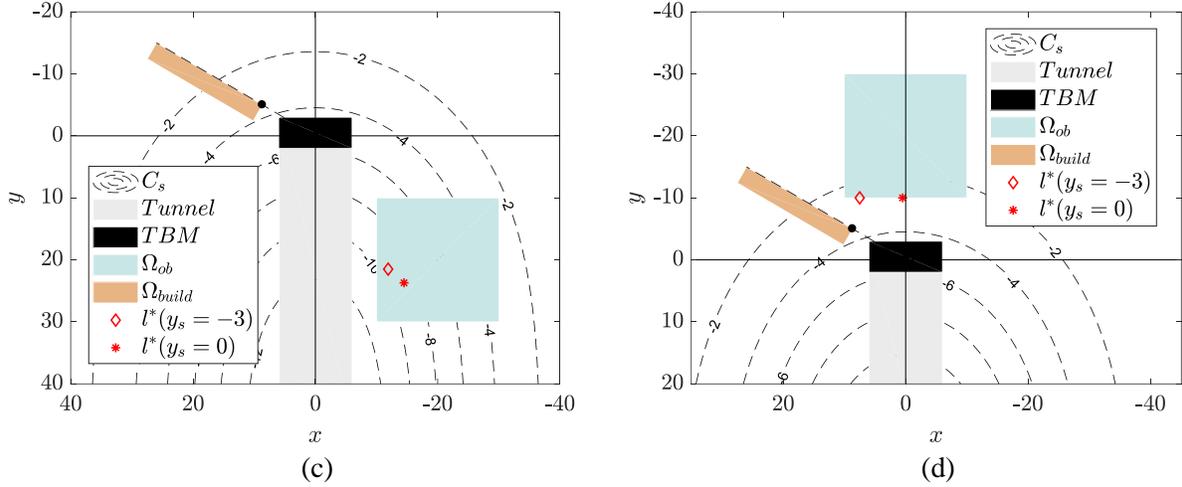

**Fig 11.** Illustration of identified $l^*$s with (a) $\Omega_{ob} = \Omega_A$, $y_s = 0$ and $-3$; (b) $\Omega_{ob} = \Omega_B$, $y_s = 0$ and $-3$ (c) $\Omega_{ob} = \Omega_C$, $y_s = 0$ and $-3$ and (d) $\Omega_{ob} = \Omega_D$, $y_s = 0$ and $-3$.

**Table 6.** Identification of $l^*$ with corresponding SOI based on different combinations of $\Omega_{ob}$ and $y_s$

| Region | Parameters (m) | $y_s$(m) | $l^*$(m) | SOI(-) |
|---|---|---|---|---|
| $\Omega_A$ | [10,30; 10,30;] | 0 | (14.7, 24.2, 0) | 0.0513 |
|  |  | -3 | (12.5, 20.2, 0) | 0.1435 |
| $\Omega_B$ | [−30, −10; −30, −10;] | 0 | (−10, −10, 0) | 0.0069 |
|  |  | -3 | (−10, −10, 0) | 0.0163 |
| $\Omega_C$ | [−30, −10; 10,30;] | 0 | (−14.5, 23.7, 0) | 0.0498 |
|  |  | -3 | (−11.9, 21.5, 0) | 0.1511 |
| $\Omega_D$ | [−10,10; −30, −10;] | 0 | (0.5, −10, 0) | 0.0133 |
|  |  | -3 | (7.5, −10, 0) | 0.0052 |

Therefore, the procedures represented above showcase a systematic approach for localizing the optimal monitoring topology for the risk assessment and tracking of a tunneling-induced structural failure. Instead of focusing on the location where the largest deformation happens, this paper sheds light on utilizing probabilistic tools to account for the uncertainties involved. Future work can investigate the uncertainty of the soil properties via random fields modelling and consider the paradigm that can handle multiple building over the tunneling contour. Though this strategy is interpreted by a pedagogically practical case of tunneling-induced settlement to the damage of building, it can be further generalized for the application in other geotechnical structure and underground infrastructure system. It is expected that this work can be leveraged to improve the efficiency for decision-making of structural health/risk monitoring of geo-structures.

## 5. Conclusions

This paper proposes a computational framework based on a novel metric called SOI (sensitivity of information) to determine the optimal monitoring location for risk tracking of geotechnical structures and underground infrastructure systems. This goal is achieved by quantitively defining the contribution of each possible monitoring location to the change in the updated reliability index. Specifically, SOI is defined as the expected changing ratio between the unconditional and conditional reliability indexes. The monitoring location that maximizes SOI is deemed to be the optimal one. Moreover, surrogate modeling is integrated to avert the otherwise very high computational cost of the optimization problem. To explore the performance of the proposed computational framework, a practical case that investigates the risk posed by tunneling-induced settlement to building damage is studied. Simulation results showcase that the optimal settlement monitoring grounded in reliability updating theory can be accurately determined. Despite the gained computational efficiency, future work can focus on further improving the computational efficiency for cases involving complex high-fidelity geotechnical/underground structural models.

**Acknowledgements**
This research was supported in part by the National Science Foundation of China (NSFC) through awards 51778337, 51890901 and 52008155, the U.S. National Science Foundation (NSF) through award CMMI-2000156, 'Shuimu Tsinghua Scholar' Plan by Tsinghua University through award 2020SM006 and Chinese Postdoctoral International Exchange Program through award YJ20210126. Any opinions, findings, and conclusions or recommendations expressed in this paper are those of the authors and do not necessarily reflect the views of these supports.